\documentclass[%
aip,
preprint,
showpacs,
preprintnumbers,
 amsmath,
 amssymb,
]{revtex4-1}

\usepackage{graphicx}
\usepackage{dcolumn}
\usepackage{bm}

\begin{document}

\title{Ion beam bunching via phase rotation in cascading laser-driven ion acceleration}

\author{H. C. Wang}
\affiliation{Key Laboratory for Laser Plasmas (MoE), School of Physics and Astronomy,
Shanghai Jiao Tong University, Shanghai 200240, China}%
\affiliation{Collaborative Innovation Center of IFSA, Shanghai Jiao Tong University,
Shanghai 200240, China}%

\author{S. M. Weng}\email{wengsuming@gmail.com}%
\affiliation{Key Laboratory for Laser Plasmas (MoE), School of Physics and Astronomy,
Shanghai Jiao Tong University, Shanghai 200240, China}%
\affiliation{Collaborative Innovation Center of IFSA, Shanghai Jiao Tong University,
Shanghai 200240, China}%

\author{M. Liu}
\affiliation{Key Laboratory for Laser Plasmas (MoE), School of Physics and Astronomy,
Shanghai Jiao Tong University, Shanghai 200240, China}%
\affiliation{Collaborative Innovation Center of IFSA, Shanghai Jiao Tong University,
Shanghai 200240, China}%

\author{M. Chen}
\affiliation{Key Laboratory for Laser Plasmas (MoE), School of Physics and Astronomy,
Shanghai Jiao Tong University, Shanghai 200240, China}%
\affiliation{Collaborative Innovation Center of IFSA, Shanghai Jiao Tong University,
Shanghai 200240, China}%

\author{M. Q. He}
\affiliation{Institute of Applied Physics and Computational Mathematics, Beijing 100094, China}

\author{Q. Zhao}
\affiliation{Key Laboratory for Laser Plasmas (MoE), School of Physics and Astronomy,
Shanghai Jiao Tong University, Shanghai 200240, China}%
\affiliation{Collaborative Innovation Center of IFSA, Shanghai Jiao Tong University,
Shanghai 200240, China}%

\author{M. Murakami}
\affiliation{Institute of Laser Engineering, Osaka University, Osaka 565-0871, Japan}

\author{Z. M. Sheng}\email{z.sheng@strath.ac.uk}%
\affiliation{Key Laboratory for Laser Plasmas (MoE), School of Physics and Astronomy,
Shanghai Jiao Tong University, Shanghai 200240, China}%
\affiliation{Collaborative Innovation Center of IFSA, Shanghai Jiao Tong University,
Shanghai 200240, China}%
\affiliation{Tsung-Dao Lee Institute, Shanghai 200240, China}
\affiliation{SUPA, Department of Physics, University of Strathclyde, Glasgow G4 0NG, United Kingdom}

\date{\today}
\begin{abstract}
The ion beam bunching in a cascaded target normal sheath acceleration is investigated by theoretical analysis and particle-in-cell simulations.
It is found that a proton beam can be accelerated and bunched simultaneously by injecting it into the rising sheath field at the rear side of a laser-irradiated foil target.
In the rising sheath field, the ion phase rotation may take place since the back-end protons of the beam feels a stronger field than the front-end protons.
Consequently, the injected proton beam can be compressed in the longitudinal direction.
At last, the vital role of the ion beam bunching is illustrated by the integrated simulations of two successive stages in a cascaded acceleration.
\end{abstract}




\maketitle

\section{Introduction}

With the advent of compact intense lasers, the generation of energetic particles in the laser-plasma interactions has been drawing an enormous amount of attention \cite{Tajima1979,Esarey2009,DaidoRPP2012,MacchiRMP2013}.
Paramountly, laser-driven plasma-based accelerators can sustain an accelerating field as strong as a few hundreds of GV/m, which is many orders of magnitude stronger than that in the conventional radio-frequency (RF) accelerators.
By such a strong accelerating field, charged particles can be accelerated to high energies over an extremely short distance.
This allows laser-driven accelerators to be more compact and more economical than the conventional RF accelerators.
Nevertheless, the particle energies achieved by laser-driven accelerators are still much lower than those by the conventional RF accelerators. In particular, it is still a big challenge to increase the ion energies in laser-driven ion acceleration. So there is still a substantial gap between the achievable ion energies of laser-driven ion sources and the necessary energies for some important applications, such as proton therapy \cite{Bulanov2002,Medphys2008}.

In order to enhance the ion energies as well as the beam quality and energy conversion efficiency in laser-driven ion acceleration, a number of mechanisms have been proposed with various laser and target parameters.
So far, the target normal sheath acceleration (TNSA) is still the most studied mechanism in experiments because of its relatively simple requirements on the laser and target conditions.
In this mechanism, a strong charge-separation electrostatic field is formed at the rear sheath of a thin target when a portion of hot electrons produced in laser-plasma interaction go through the target \cite{WilksPOP2001}. By this sheath field, the ions can be accelerated up to tens MeV per nucleon \cite{WagnerPRL2016}.
Alternatively, the generation of energetic ions by the radiation pressure acceleration (RPA) has been intensively studied recently \cite{RPApapers1}.
According to the theoretical model, the ions of an ultrathin foil can be steadily accelerated by the radiation pressure of a circularly-polarized intense laser pulse. Consequently, it is predicted that the ions could be accelerated by the RPA to much higher energies with a high conversion efficiency.
However, the RPA has some extremely stringent requirements on the laser and target conditions in experiments \cite{BinPRL2015,KimPoP2016}, including an ultrathin foil and an ultrahigh contrast laser pulse. Furthermore, the laser pulse should be circularly-polarized for suppressing the plasma heating, which is also a big challenge for the high-power laser pulse \cite{WengOptica}.

In addition to individual acceleration schemes, some hybrid schemes \cite{LLYu,Liu2018,Higginson} or cascaded schemes \cite{ZhengJ,ChenM,Pfotenhauer,WangWP,WangPRE2014,Kawata2014,Pei,Kawata2016,WangPOP2017} have also been proposed for enhancing the ion energies in laser-driven ion acceleration.
Recently, the efficient generation of protons with energies exceeding 94 MeV by a combination of the RPA and TNSA mechanisms has been demonstrated in the experiment \cite{Higginson}.
More interestingly, some cascaded acceleration schemes allow the spectral shaping of the resulting ion beams \cite{Pfotenhauer}. As a result, the ion energies and the energy spread could be simultaneously improved in a cascaded acceleration scheme \cite{ZhengJ,WangWP}.
However, the improvement on the energy spread will disappear if the ion beam duration is comparable to the lifetime of the accelerating field \cite{WangWP}. Therefore, the bunching would be critically required to control the longitudinal size of the ion beam in a cascaded laser-driven acceleration scheme \cite{Kawata2016}.

\begin{figure}
\includegraphics[width=0.48\textwidth]{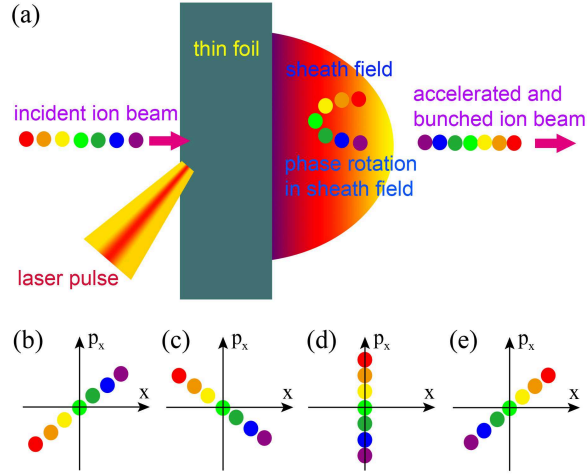}
\caption{ (color online).
Schematic of a cascaded TNSA, (a) A strong electrostatic sheath field is generated at the rear side of a thin foil target that is irradiated by a short intense laser pulse.
An incident ion beam can be accelerated and well bunched if it is injected into during the rising stage of the sheath field.
Distributions of injected ions in the $x-p_x$ phase space (b) before the acceleration, (c) and (d) during the acceleration, and (e) after the acceleration. The coordinate origins of the $x-p_x$ phase space are modified simultaneously with the moving and accelerating of the ion beam center. A half-cycle rotation of these ions in the phase space is accomplished in a single acceleration stage.
}\label{scheme}
\end{figure}

In this paper, we propose a cascaded TNSA scheme, in which the ion beam bunching is realized via the ion phase rotation in the sheath field at the target rear side.
As a typical configuration for the well-known TNSA mechanism, a thin foil target is irradiated by a short intense laser pulse in Fig. \ref{scheme}(a).
A large number of hot electrons will be generated via collisionless heating mechanisms \cite{ShengCPB,Mulser}. A strong charge-separation electrostatic field can be stimulated at the rear sheath when some hot electrons propagate through the target and expand into the vacuum  \cite{Mora2003}.
In particular, it takes a considerable period of time before this sheath field attains its maximum strength \cite{WangWP}.
If an ion beam is injected into this laser-irradiated foil when its sheath field is rising, one can imagine that the back-end ions of the beam will feel a stronger sheath field than the front-end ions.
Therefore, the back-end ions may be accelerated to higher velocities even if their initial velocities are lower than those of the front-end ions as shown in Figs. \ref{scheme}(b) and (c).
Consequently, the back-end ions will overtake the front-end ions as shown in Figs. \ref{scheme}(d) and (e).
Concerning this whole process, a half-cycle rotation of the ions in the phase space is accomplished.
Using particle-in-cell (PIC) simulations, we verify that the longitudinal size of the injected ion beam can be well controlled via such a phase rotation, i.e., the ion beam bunching is achieved.
Furthermore, the importance of the ion beam bunching in a cascaded acceleration is clarified by the integrated simulations of two successive acceleration stages.



\section{Simulation results and analysis}

To visualize the ion beam bunching via the phase rotation, we have performed a series of 2D3V particle-in-cell (PIC) simulations of cascaded TNSA using the code EPOCH \cite{ArberEPOCH}.

In the first set of simulations, a single stage of cascaded TNSA is investigated. In the simulations, a linearly-polarized laser pulse irradiates a carbon target obliquely with an angle of $45^\circ$ from the left side.
It is assumed that the laser pulse has a wavelength $\lambda=1 \mu m$, a peak intensity $I_0 \simeq 3.08\times10^{20}$ W/cm$^2$ (the normalized vector potential $a_0\equiv|e\textbf{E}/\omega m_e c|=15$).
The pulse has Gaussian intensity profiles in both the transverse and longitudinal directions with a spot radius $\sigma=5\mu m$ and a duration $\tau$.
In a typical simulation, we set the duration $\tau = 12T_0$, where $T_0=2\pi/\omega$ is laser wave period.
The time evolution of the sheath field are compared among the cases with different durations $\tau$=6, 12, and 18$T_0$.
The simulation box has a size of $80\lambda \times 20\lambda$, the spatial resolutions are $\Delta x=\Delta y=\lambda/100$, and 50 macro-particles per species per cell are allocated in the target region.
The target is assumed to be a uniform fully-ionized Carbon foil with $\rho \approx 0.057$ g/cm$^3$ (the electron number density $n_e=15 n_c$) that locates in $0 \leq x/\lambda \leq 6 $ and $-10 \leq y/\lambda \leq 10$.
Here, a near-critical-density target is used instead of a solid target in order to enhance the coupling of laser energy into hot electrons and hence the ion acceleration \cite{Higginson,BinPRL2015,WengNJP}.
In addition, a quasi-monoenergetic proton beam is injected along the $x$-axis into the laser-irradiated foil.
The proton beam initially has a size of $0.2\times0.6\mu m^2$ and a density $n_p=0.1n_c$. The mean proton energy is assumed to be a function of $x$ as $E_0(x)=[10 + 10(x-x_0)/\lambda]$ MeV for $|x-x_0|<0.1 \lambda$, where $(x_0,0)$ is the coordinates of the beam center. Correspondingly, the mean velocity of the protons at the beam center is about $v_{x0}\simeq 0.145c$.
Although the mean proton energy is a function of $x$, an initially uniform temperature $T_i = 1$ keV is assumed.
By changing the initial center $x$-coordinate $x_0$, the time when the proton beam arrives at the target rear ($x=6\lambda$) and enters into the sheath field can be well controlled in simulations.
For reference, the peak of the laser pulse is assumed to arrive at the target front ($x=0$) at $t=0$, and the simulations begin at $t \simeq -16T_0$.

\subsection{Mechanism of ion beam bunching}

\begin{figure}
\includegraphics[width=0.48\textwidth]{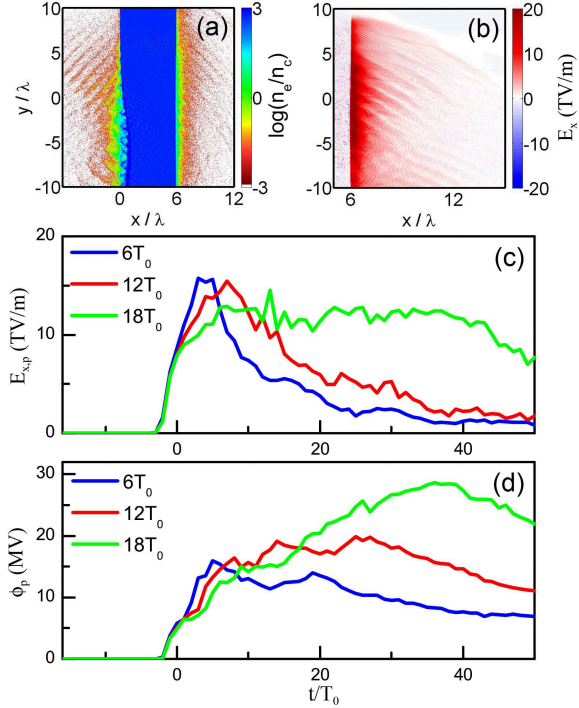}
\caption{ (color online).
Spatial distributions of (a) the electrons and (b) the sheath field $E_x$ at $t=10T_0$ in the simulation case with a laser pulse duration $\tau=12T_0$.
Time evolutions of (c) the sheath field peaks and (d) the electrostatic potential peaks in the simulation cases with different laser pulse durations $\tau=6$, 12, and 18$T_0$.
}\label{figDynamics}
\end{figure}

To understand the mechanism of the ion beam bunching in a cascaded TNSA, we first revisit the generation of hot electrons and  electrostatic sheath field in a laser-foil interaction.
Thanks to the efficient coupling of laser energy into the applied near-critical-density target, a large number of hot electrons are generated. As shown in Fig. \ref{figDynamics}(a), a considerable portion of these hot electrons can penetrate through the target and expand into the vacuum at the rear side.
As a result, an electrostatic field as strong as a few TV/m is quickly rising in the charge-separation sheath at the rear side as shown in Fig. \ref{figDynamics}(b).
Figure \ref{figDynamics}(c) displays the time evolutions of the sheath field peaks $E_{x,p}$ with different laser pulse durations. Correspondingly, the time evolutions of the electrostatic potential peaks $\phi_p$ are displayed in Fig. \ref{figDynamics}(d).
It is illustrated that the rising of the sheath field needs a response time on the order of the laser pulse duration $\tau$ if the latter is around ten laser wave periods.
In a typical simulation case with $\tau = 12T_0$, the sheath field peak rises from 0 to the maximum $\sim 15$ TV/m in a time interval $-4 T_0\leq t \leq 8 T_0$.
The quick rising of the sheath field, usually accompanied by the broadening of the charge-separation sheath, will greatly raise the electrostatic potential peak $\phi_p$ as shown in Fig. \ref{figDynamics}(d).
Because of the broadening of the charge-separation sheath, the electrostatic potential peak $\phi_p$ can continuously increase for some time after the sheath field peak achieves its maximum.

Figures \ref{figDynamics}(c) and \ref{figDynamics}(d) indicate that the back-end protons of the beam can obtain more energy from a stronger sheath field than the front-end protons if the proton beam arrives during the rising stage of the sheath field.
The obtained energy of an proton from the TNSA is roughly approximate to the electrostatic potential peak $\phi_p$ when this proton enters into the sheath field.
For a quasi-monoenergetic proton beam of a length of $L_0$ locating in the region $|x-x_0| \leq L_0/2$, we assume that the mean proton energies (velocities) at the front-end, the center and the back-end are $E_f$ ($v_f$), $E_c$ ($v_c$) and $E_b$ ($v_b$), respectively. Usually $E_f>E_c>E_b$ and $v_f>v_c>v_b$ are satisfied since the protons with higher energies will naturally propagate at the front after a considerable long propagation.
Consequently, the ion rotation in the phase space can take place, i.e., the back-end protons overtake the front-end protons, only under the condition
\begin{equation}
\phi_p(t_b)-\phi_p(t_f) > E_f - E_b,
\end{equation}
where $t_f=(t_0 v_c - L_0/2)/v_f$, $t_b=(t_0 v_c +L_0/2)/v_b$, and $t_0$ are the times when the front-end, the back-end, and the center protons enter into the sheath field, respectively.

\begin{figure}
\includegraphics[width=0.48\textwidth]{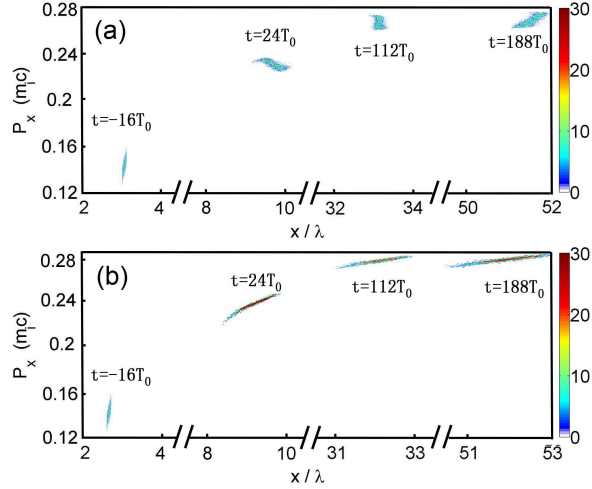}
\caption{ (color online).
Time evolutions of the proton distributions in the $x-p_x$ phase space in a single acceleration stage of a cascaded TNSA for two different proton beams with initial center $x$-coordinates (a) $x_0=3.2\lambda$, and (b) $x_0=2.6\lambda$, respectively.
In the simulations, the laser pulse duration is $\tau=12 T_0$.
The initial moment $t=-16T_0$ is before the acceleration, $t=24T_0$ and $112T_0$ are during the acceleration, and $t=188T_0$ is after the acceleration.
}\label{protonphase}
\end{figure}

While under the condition $\phi_p(t_b)-\phi_p(t_f) \sim E_f - E_b$, a highly monoenergetic proton beam can be expected \cite{ZhengJ,WangWP}. In order to achieve the ion beam bunching during a single accelerate stage, however, the difference in the electrostatic potential $|\phi_p(t_b)-\phi_p(t_f)|$ must be substantially larger than the initial difference in the mean proton energy $|E_f - E_b|$. On the other hand, the final energy spread of the obtained proton beam will be obviously increased if the potential difference $|\phi_p(t_b)-\phi_p(t_f)|$ is much larger than $|E_f - E_b|$.
Considering the above two aspects, $\phi_p(t_b)-\phi_p(t_f) \simeq 2(E_f - E_b)$ should be a relatively appropriate condition for achieving the efficient ion beam bunching and controlling the final energy spread on the initial level.

\subsection{Results of a single acceleration stage}

\begin{figure}
\includegraphics[width=0.48\textwidth]{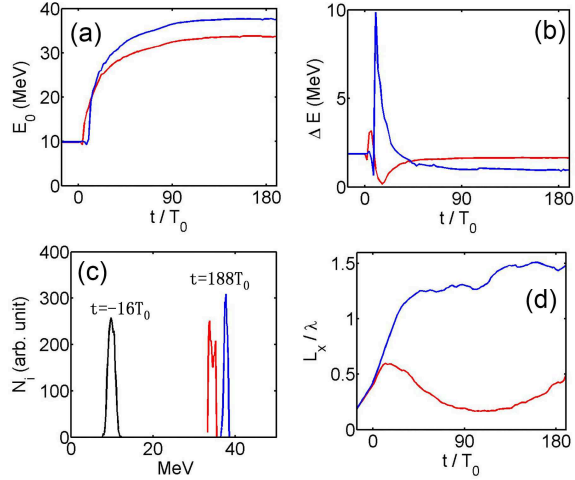}
\caption{ (color online).
Time evolutions of (a) the mean energy $E_0$, (b) the absolute FWHM energy spread ($\Delta E$), and (d) the FWHM longitudinal size ($L_x$) in a single acceleration stage for two different injected proton beams initially centered at $x_0=3.2\lambda$ (red solid curves) and $2.6\lambda$ (blue solid curves), respectively.
The corresponding final energy spectra at $t=188T_0$ are compared in (c), where their initial energy spectra are same and displayed by the black curve.
}\label{qualities}
\end{figure}

To some extent, the electrostatic potential difference $|\phi_p(t_b)-\phi_p(t_f)|$ can be controlled by modifying the arrival time of the proton beam at the sheath field, while the latter is realized by changing the initial $x$-coordinate $x_0$ of the proton beam center in the simulations.
Figure \ref{protonphase} compares the time evolutions of the proton distributions in the $x-p_x$ phase space for two different injected proton beams initially centered at $x_0=3.2\lambda$ and $2.6\lambda$, respectively.
In the case $x_0=3.2\lambda$, the front-end and back-end protons will arrive at the target rear side at $t_f=1.8T_0$ and $t_b=5.1T_0$, respectively.
As shown in Fig. \ref{figDynamics}(d), the corresponding electrostatic potential difference $\phi_p(t_b)-\phi_p(t_f) \simeq 5$ MeV, which is about the double of the initial proton energy difference between the front-end and back-end protons.
Consequently, the back-end protons will obtain more energy and quickly become faster than the front-end protons as indicated by the phase space distribution at $t=24T_0$ in Fig. \ref{protonphase}(a).
Subsequently, the back-end protons can catch up with the front-end protons and the whole proton beam is highly compressed at $t=112T_0$.
After this, the length of the proton beam will gradually increase due to the free expansion, and the proton distribution in the $x-p_x$ phase space becomes similar to the initial one.
In this whole stage, a half-cycle rotation of these protons in the $x-p_x$ phase space is evidenced.
In contrast, the proton beam will enter into the sheath field relatively later with $t_f=5.7T_0$ and $t_b=9.5T_0$ in the case $x_0=2.6\lambda$. Correspondingly, the electrostatic potential difference is reduced to $\phi_p(t_b)-\phi_p(t_f) \simeq 1.5$ MeV, which is slightly smaller than the initial proton energy difference.
As a result, the back-end protons will be accelerated to be nearly as fast as the front-end protons as shown in Fig. \ref{protonphase}(b). However, the length of the proton beam always increases in the whole stage.

\begin{figure}
\includegraphics[width=0.48\textwidth]{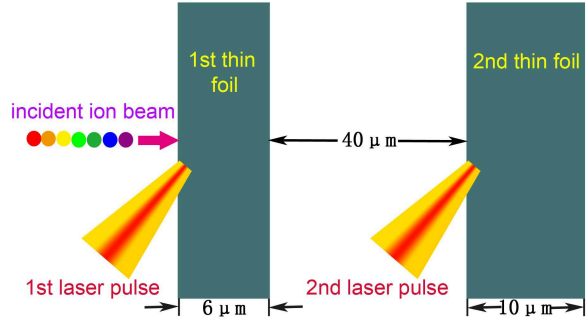}
\caption{ (color online).
Configuration for the integrated simulations of two successive acceleration stages.
In each stage, a linearly-polarized laser pulse irradiates a carbon foil target obliquely with an angle of $45^\circ$.
The thicknesses of the first and the second targets are 6 $\mu$m and 10 $\mu$m, respectively. The distance between two foil targets is 40 $\mu$m.
The peak of the first laser pulse is assumed to arrive at the first target front ($x=0$) at $t=0$.
While the second laser pulse peak arrives at the front of the second target ($x=46$) with a time delay of $t=194 T_0$ or $175T_0$, respectively.
In addition, a quasi-monoenergetic proton beam, centered at $x_0=3.2\lambda$ or $2.6\lambda$, is incident along the $x$-axis into laser-irradiated foil targets.
}\label{SchemeTS}
\end{figure}

The time evolutions of the beam qualities in a single acceleration stage are quantitatively compared between the cases with $x_0=3.2\lambda$ and $2.6\lambda$ in Fig. \ref{qualities}.
Figure \ref{qualities}(a) indicates that the proton beam initially centered at $x_0=3.2\lambda$ can be accelerated up to a mean energy of $\sim34$ MeV which is slightly lower than that of the one initially at $x_0=2.6\lambda$, since the latter enters the rising sheath field later and feels a stronger field.
Further, Fig. \ref{qualities}(b) demonstrates that the FWHM absolute energy spread can be efficiently suppressed to $\sim 1.2$ MeV in the $x_0=2.6\lambda$ case while it remains at around the initial value $\sim2$ MeV in the $x_0=3.2\lambda$ case.
As a result, Fig. \ref{qualities}(c) shows that the energy spectrum in the $x_0=2.6\lambda$ case seems superior to the one in the $x_0=3.2\lambda$ case.
However, it must be pointed out that in the $x_0=2.6\lambda$ case the FWHM longitudinal size of the proton beam monotonously increases up to $\sim 1.5\lambda$, which is about one order of magnitude higher than the initial length.
While the proton beam experiences an obvious shrink in the $x_0=3.2\lambda$ case, the minimum FWHM longitudinal can be even a bit smaller than the initial size. This implies that the ion beam bunching is achieved simultaneously with the acceleration in the latter case. In the following paragraphs, we will show the importance of such ion beam bunching in the cascaded acceleration.


\subsection{Results of two successive acceleration stages}

\begin{figure}
\includegraphics[width=0.48\textwidth]{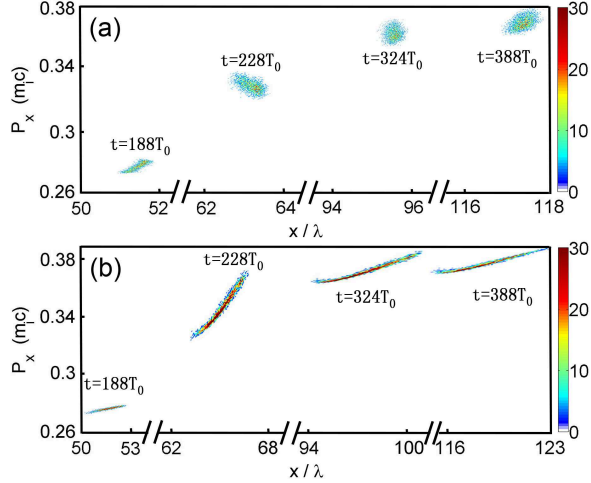}
\caption{ (color online).
Time evolutions of the proton distributions in the $x-p_x$ phase space in the second acceleration stage of a cascaded TNSA for two different proton beams with initial center $x$-coordinates (a) $x_0=3.2\lambda$, and (b) $x_0=2.6\lambda$, respectively.
The moment $t=188T_0$ is after the first acceleration but before the second acceleration, $t=228T_0$ and $324T_0$ are during the second acceleration, and $t=388T_0$ is after the second acceleration.
}\label{phaseTS}
\end{figure}

To illuminate the crucial role of the ion beam bunching in the cascaded acceleration, the integrated simulations of two successive acceleration stages are performed. As shown in Fig. \ref{SchemeTS}, the laser pulse and the foil target in the first stage are exactly the same ones used in the above single acceleration stage. While the laser pulse and the foil target in the second stage are similar to those in the first stage, except that the second foil thickness increases to 10 $\mu$m in order to prevent the direct exposure of the proton beam into the laser field. The distance between two foil targets is 40 $\mu$m.
Then the proton beam initially centered at $x_0=3.2\lambda$, after the first acceleration, will arrive at the rear side of the second target in the time interval $202 T_0 \leq t \leq 204 T_0$.
In this case the peak of the second pulse is set to arrive at the front of the second target at $t\simeq 194 T_0$,
hence the proton beam will enter into the second sheath field when it is quickly rising as in the first stage.
As a result, in the second acceleration stage this proton beam can experience another half-cycle phase rotation as shown in Fig. \ref{phaseTS}(a).
While the proton beam initially centered at $x_0=2.6\lambda$, after the first acceleration, will arrive at the rear side of the second target in the time interval $196 T_0 \leq t \leq 204 T_0$.
Then the peak of the second pulse is set to arrive at the front of the second target at $t\simeq 175 T_0$, so this proton beam will arrive at around the time of the second sheath field peak as in the first stage.
Consequently, Fig. \ref{phaseTS}(b) shows no phase rotation in this case.

\begin{figure}
\includegraphics[width=0.48\textwidth]{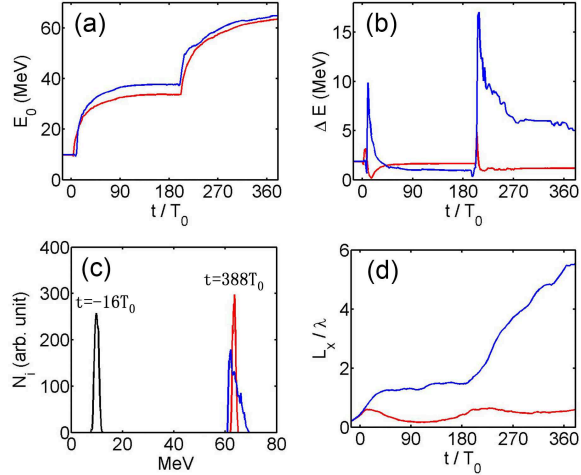}
\caption{ (color online).
Time evolutions of (a) the mean energy $E_0$, (b) the absolute FWHM energy spread ($\Delta E$), and (d) the FWHM longitudinal size ($L_x$) in two successive acceleration stages for two different injected proton beams initially centered at $x_0=3.2\lambda$ (red solid curves) and $2.6\lambda$ (blue solid curves), respectively.
The corresponding final energy spectra at $t=388T_0$ are compared in (c), where their initial energy spectra are same and displayed by the black curve.
}\label{qualitiesTS}
\end{figure}

The time evolutions of the beam qualities during these two successive acceleration stages in the cases with and without the bunching are compared in Fig. \ref{qualitiesTS}.
Figure \ref{qualitiesTS}(a) shows that the energy of an initial 10 MeV proton beam can be boosted up to more than 60 MeV after two successive acceleration stages regardless of the ion beam bunching.
As explained above, the scheme with the bunching is not the optimal choice for suppressing the energy spread in a single acceleration stage.
However, it is worth noting that the energy spread can be well controlled after two acceleration stages in the case with the bunching as shown in Fig. \ref{qualitiesTS}(b).
While the energy spread in the case without the bunching is increased obviously after the second acceleration stage.
This is because without the bunching the proton beam will be greatly prolonged in the first acceleration stage, and it is much harder to control the energy spread of this prolonged proton beam in the second stage.
As a result, the final energy spectrum in the case with the bunching can surpass the one without the bunching as displayed in Fig. \ref{qualitiesTS}(c).
More importantly, Fig. \ref{qualitiesTS}(d) shows that without the bunching the FWHM longitudinal size of the proton beam will dramatically increase up to $\sim 5.5\lambda$, which is severely adverse to the next acceleration stages.
In contrast, the longitudinal size of the proton beam can be kept on the level of the initial size with the bunching.
Therefore, with the bunching the acceleration of such a high-quality proton beam can be repeated continuously in the next acceleration stages.

\section{Discussion and Conclusion}

It is worth pointing out that the ion rotation in the phase space could also take place after two or more acceleration stages in the cases that the ratio between the electrostatic potential difference and the initial mean energy difference $[\phi_p(t_b)-\phi_p(t_f)]/(E_f - E_b) < 1$.
However, the longitudinal size of the ion beam in these cases may have already increased up to an intolerable level after two or more acceleration stages.
Therefore, it would be better to achieve the efficient ion beam bunching in each single stage under the condition that the electrostatic potential difference is substantially large enough to compensate for the mean energy difference, i.e., $[\phi_p(t_b)-\phi_p(t_f)]>(E_f - E_b)$.

In order to save the computational cost, the laser-target parameters in the simulations are set to guarantee $[\phi_p(t_b)-\phi_p(t_f)]/(E_f - E_b) \sim 2$.  This allows the ion rotation in the phase space take place as soon as possible, in the meanwhile the beam energy spread doesn't obviously increase.
In the cases that $[\phi_p(t_b)-\phi_p(t_f)]$ is slightly larger than $(E_f - E_b)$, the ion rotation in the phase space and the energy spread reduction would be achieved simultaneously in a single acceleration stage. This implies that not only the longitudinal size but also the energy spread of an injected ion beam could be effectively suppressed in a cascaded laser-driven ion acceleration scheme.
More importantly, the ion rotation in the phase space will take place at a much slower pace in these cases. Accordingly, the distance between two successive targets should be enlarged dramatically.
Although a much larger distance between two targets may cause inconvenience to the PIC simulations, it could be more favorable to the experimental realizations.

Besides the laser pulse and the foil target as employed in a normal TNSA scheme, an initially quasi-monoenergetic proton beam is crucially required in this cascaded laser-driven ion acceleration scheme.
Such a proton beam might be obtained from the TNSA with energy selection \cite{ToncianScience2006} or a laser-driven nanotube accelerator \cite{MurakamiAPL2013} and so on.

In summary, we find that the ion beam bunching in a cascaded TNSA scheme can be achieved via the ion rotation in the phase space.
By modifying the time delay between the injected proton beam and the laser pulse, one can allow the proton beam enter into the sheath field at the target rear side when it is quickly rising. Then the back-end protons of the beam will feel a stronger sheath field and be accelerated to higher energies than the front-end protons. Consequently, the back-end protons will overtake the front-end protons, i.e., the ion phase rotation takes place.
More importantly, the integrated simulations of two successive acceleration stages verifies that the energy spread in a cascaded acceleration can be well controlled only when the ion beam is bunched via such phase rotations.

\begin{acknowledgments}
The work was supported by the National Basic Research Program of China (Grant No. 2013CBA01504), National Natural Science Foundation of China (Grant Nos. 11675108, 11655002, 11721091, and 11535001), National 1000 Youth Talent Project of China, Science and Technology Commission of Shanghai Municipality (Grant No. 16DZ2260200).
Simulations have been carried out on the Pi supercomputer at Shanghai Jiao Tong University.
\end{acknowledgments}


\bibliography{apssamp}

\end{document}